\documentstyle[aps,12pt,epsf]{revtex}

\newcommand\pubnumberk{KA--TP--23--2001}
\newcommand\pubdate{\today}

\newcommand\hepnumber{hep-ph/0109103}

\def\csumb{
Institut f\"ur Theoretische Physik,
Universit\"at Karlsruhe, \\ D-76128 Karlsruhe, Germany}

\def\Title#1{\begin{center} {\Large\bf #1 } \end{center}}
\def\Author#1{\begin{center}{ \sc #1} \end{center}}
\def\Address#1{\begin{center}{ \it #1} \end{center}}

\newcommand\pubblock{\rightline{\begin{tabular}{l} 
                                                   \pubnumberk\\
         \pubdate\\ \hepnumber \end{tabular}}}
\newenvironment{Abstract}{\begin{quotation}  }{\end{quotation}}

\newcommand{\hpm}{$H^{\pm}$}
\newcommand{\msbar}{$\overline{MS}$}

\begin{document}
\begin{titlepage}
\pubblock

\Title{ $O(\alpha_s)$ Corrections to $b \bar b \rightarrow
W^{\pm}H^{\mp}$ at the CERN Large Hadron Collider}
\vfill
\Author{  Wolfgang Hollik and  Shou-hua Zhu }
\Address{\csumb}
\vfill
\begin{Abstract} 
The $O(\alpha_{s})$ corrections to the cross section for $b \bar b \rightarrow
W^{\pm}H^{\mp}$ at the
LHC are calculated in the minimal supersymmetric standard model (MSSM) in 
the $\overline{MS}$ and OS (on-mass-shell) renormalization
schemes. The results in two schemes are in good agreement.
In the $\overline{MS}$
scheme, 
the QCD corrections are negative and within $ -14\% \sim -20\%$ for charged
Higgs mass up to 1 TeV and
$\tan\beta >15$.
For $\tan\beta =2$,
the magnitude of the QCD corrections can be greater than
$30\%$.

\end{Abstract}
\vfill


\end{titlepage}

\eject \baselineskip=0.3in
\section{ Introduction}

The detection of Higgs particles is 
one of the most important objectives of the Large Hadron
Collider (LHC).
Charged Higgs bosons 
are predicted in extended versions of the Standard model (SM),
like two-Higgs-doublet models (2HDM) and 
the Minimal Supersymmetric Standard Model (MSSM).
A discovery of such an additional Higgs boson will
immediately indicate physics beyond the SM; there is, hence, 
strong theoretical and experimental activity
to provide the basis for its accurate exploration.

At hadron colliders, the charged Higgs boson $H^\pm$
could appear as  
the decay product of primarily produced top quarks
if the mass of $H^\pm$ is smaller than $m_t - m_b$.
For heavier $H^\pm$, other mechanism 
for $H^\pm$ production have been investigated: 
single Higgs-boson production
associated with heavy quarks, like
$g b \to H^- t$\cite{Gbht}
and $q b \to q' b H^-$\cite{Qbhqb}, and
pair production of $H^\pm$
through tree-level $q \overline{q}$ 
annihilation 
and via the loop-induced gluon-fusion 
mechanism~\cite{All1}.
Moreover,
single \hpm production in association with a $W$ boson,
via tree-level $b\bar b$ annihilation 
and one-loop $gg$ fusion has been proposed
and analyzed~\cite{Dicus}.
Detailed  studies \cite{All2} show that these production
mechanisms at the LHC can help to explore the 
parameter space, even beyond $m_{H^{\pm}}\sim 1$TeV and down
to at least $\tan\beta \sim 3$.

Since $b\bar b$ annihilation is the main source
of the hadronic $W^{\pm}H^{\mp}$ production cross section,
it is necessary to calculate and implement 
also the loop contributions to $b\bar{b} \rightarrow W^{\pm}H^{\mp}$
for more accurate theoretical predictions.
Recently, 
the calculation of the
$O(\alpha_{ew}m_{t(b)}^{2}/m_{W}^{2})$ and
$O(\alpha_{ew}m_{t(b)}^{4}/m_{W}^{4})$ supersymmetric 
electroweak(EW) one-loop corrections 
were presented in~\cite{Yang} 
in the frame of the MSSM,
which can give rise to a 10-15\% reduction of the lowest-order result.
In this paper, we deal with the one-loop QCD corrections to
$b\bar b \rightarrow W^{\pm}H^{\mp}$.

The arrangement of this paper is as follows. Section~II contains the
analytic results, and in Section~III  we present numerical examples
and discuss the implications of our results 
The lengthy expressions of the form factors are
collected in the Appendix.

\section{ Analytic expressions}

\subsection{Virtual corrections}
The Feynman diagrams for charged Higgs-boson production via
the parton process
$b(p_1)\bar b(p_2)\rightarrow H^-(k_1) W^{+}(k_2)$,
including the QCD corrections, are shown in Fig.~1.
The diagrams are created by use of 
FeynArts~\cite{Feynarts} and are handled with the help of   
FeynCalc~\cite{FeynCalc}.
We keep the finite $b$-quark mass throughout the calculation, 
in order to control the collinear divergences.

As usual, we define the Mandelstam variables as
\begin{eqnarray}
\hat s =(p_1 +p_2)^2 =(k_1 +k_2)^2, \nonumber \\ \hat t =(p_1 -k_1)^2
=(p_2 -k_2)^2, \nonumber \\ \hat u =(p_1 -k_2)^2 =(p_2 -k_1)^2.
\end{eqnarray}

Taking into account the $O(\alpha_{s})$ corrections, the
renormalized amplitude for $b\bar b\rightarrow W^{+}H^{-}$ can be
written in the following way,
\begin{eqnarray}
\label{amplitude}
 M_{\rm ren}& =& M_0^{(s)} +M_0^{(t)} +\delta \hat M^{V_1(s)}(H)
+\delta \hat M^{V_1(s)}(h) +\delta \hat
M^{V_1(s)}(A) \nonumber \\
&&  +\delta \hat M^{V_1(t)} +\delta \hat M^{S(t)} +\delta \hat
M^{V_2(t)} +\delta M^{\rm box}.
\end{eqnarray}
$M_0^{(s)}$ and $M_0^{(t)}$ are the $s$- and $t$-channel
tree-level diagrams corresponding to
Fig.~1(a) and Fig.~1(b),  which are
given by
\begin{eqnarray}
 M_0^{(s)} &=&
  \frac{gh_b}{\sqrt{2}} 
  \left[ -\frac{\cos\alpha \sin(\alpha-\beta)}{\hat s-m_{H}^2}
+\frac{\sin\alpha \cos(\alpha-\beta)}{\hat s-m_{h}^2} \right]
  (M_5+M_6+M_9+M_{10})  \nonumber \\
&&+\frac{gh_b\sin\beta}{\sqrt{2} \, (\hat
  s-m_{A}^2)} \, (M_5-M_6+M_9-M_{10}) \, , \nonumber \\[0.2cm]
 M_0^{(t)} &=&\frac{ g}{\sqrt{2}(\hat t-m_t^2)} \,
  [h_b\sin\beta(2 M_9+m_b M_1+M_3)-h_tm_t\cos\beta M_2 ] \, ,
\end{eqnarray}
where
\begin{equation}
\label{Yukawa}
h_b =  \frac{gm_b}{\sqrt{2} m_W \cos\beta} \, , \quad 
h_t =  \frac{gm_t}{\sqrt{2}m_W\sin\beta}
\end{equation} 
denote the Yukawa couplings of the
bottom and top quarks.
$M_i$ are reduced standard matrix elements, which are
defined by
\begin{eqnarray}
\label{me}
&&M_1= \bar v(p_2) \not\varepsilon(k_2) P_R u(p_1),\nonumber
\\&&M_2= \bar v(p_2) \not\varepsilon(k_2) P_L u(p_1),\nonumber
\\&&M_{3}= \bar v(p_2) \not k_2 \not\varepsilon(k_2) P_R u(p_1),\nonumber
\\&&M_{4}= \bar v(p_2) \not k_2 \not\varepsilon(k_2) P_L u(p_1), \nonumber
\\&&M_5= \bar v(p_2) P_R u(p_1)p_1\cdot\varepsilon(k_2),\nonumber
\\&&M_6= \bar v(p_2) P_L u(p_1)p_1\cdot\varepsilon(k_2),\nonumber
\\&&M_7= \bar v(p_2) \not k_2 P_R u(p_1)p_1\cdot\varepsilon(k_2),\nonumber
\\&&M_8= \bar v(p_2) \not k_2 P_L u(p_1)p_1\cdot\varepsilon(k_2),\nonumber
\\&&M_9= \bar v(p_2) P_R u(p_1)p_2\cdot\varepsilon(k_2),\nonumber
\\&&M_{10}= \bar v(p_2) P_L u(p_1)p_2\cdot\varepsilon(k_2),\nonumber
\\&&M_{11}= \bar v(p_2) \not k_2 P_R u(p_1)p_2\cdot\varepsilon(k_2),\nonumber
\\&&M_{12}= \bar v(p_2) \not k_2 P_L u(p_1)p_2\cdot\varepsilon(k_2),
\end{eqnarray}
with the projectors $P_{L,R} \equiv (1\mp \gamma_5)/2$. 

The terms $\delta\hat{M}$ in (\ref{amplitude})
describe the virtual contributions from the 2- and 3-point functions
together with their counterterms, and $\delta M^{\rm box}$ denotes the
contribution from the irreducible 4-point function.
The vertex and
self-energy corrections to the tree-level process are included in
$\delta \hat M^{V,S}$, which are given by
\begin{eqnarray}  \delta \hat M^{V_1(s)}(H,h)
&=&
  \frac{gh_b}{\sqrt{2}} [ -\frac{\cos\alpha \sin(\alpha-\beta)}{\hat s-m_{H}^2}
,\frac{\sin\alpha \cos(\alpha-\beta)}{\hat s-m_{h}^2}]
  (M_5+M_6+M_9+M_{10}) \nonumber \\
  && \times [\frac{\delta m_b}{m_b}+\frac{1}{2}\delta Z^b_L +\frac{1}{2}\delta Z_R^b]
+\delta M^{V_1(s)}(H,h),\nonumber
\\  \delta \hat M^{V_1(s)}(A) &=&\frac{gh_b\sin\beta}{\sqrt{2}(\hat
  s -m_A^2)}[\frac{\delta m_b}{m_b}
   +\frac{1}{2}\delta Z^b_L +\frac{1}{2}\delta Z_R^b 
  ](M_5-M_6+M_9-M_{10}) +\delta M^{V_1(s)}(A),\nonumber
\\ \delta \hat M^{V_1(t)} &=&
 \frac{ g}{\sqrt{2}(\hat t-m_t^2)}
  [h_b\sin\beta(2 M_9+m_b M_1+M_3)-h_tm_t\cos\beta M_2 ] \nonumber \\
   && \times (\frac{1}{2}\delta Z^t_L +\frac{1}{2}\delta Z_L^b)
 +\delta M^{V_1(t)},\nonumber
\\  \delta \hat M^{S(t)} &=&
 \frac{ g}{\sqrt{2}(\hat t-m_t^2)^2}
  [h_b\sin\beta(2 M_9+m_b M_1+M_3) (2 m_t \delta m_t +
( m_t^2-\hat t) Z_L^t)\nonumber \\
&&  -h_t\cos\beta M_2 (
 \delta m_t ( m_t^2+\hat t)+ m_t  (m_t^2-\hat t) Z_L^t) ]
  + \delta M^{S(t)},\nonumber
\\  \delta \hat M^{V_2(t)} &=&
 \frac{ g}{\sqrt{2}(\hat t-m_t^2)}
  [h_b\sin\beta(2 M_9+m_b M_1+M_3) ( \frac{ \delta m_b}{ m_b }
+\frac{1}{ 2 } Z_R^t+\frac{1}{ 2 } Z_L^b) \nonumber \\
&&
-h_tm_t\cos\beta M_2( \frac{ \delta m_t}{ m_t }
+\frac{1}{ 2 } Z_L^t+\frac{1}{ 2 } Z_R^b) ]
  +\delta M^{V_2(t)}.
\end{eqnarray}
Therein,
$\delta M^{V_1(s)}(H,h,A)$,
$\delta M^{V_1(t)}$, $\delta M^{S(t)}$,
$\delta M^{V_2(t)}$  represent the unrenormalized
one-loop corrections 
arising, respectively, from 
the $b\bar bH(h,A)$ vertex diagrams in Fig.~1(c), 
the $\bar t bH^-$ vertex diagram Fig.~1(d),
the $\bar btW^+$ vertex diagram Fig.~1(e), and 
the top-quark self-energy diagram Fig.~1(f);
$\delta M^{\rm box}$ corresponds to
the box diagram Fig.~1(g).
All the $\delta M^{V,S}$ and $\delta M^{\rm box}$ can be written in
the form
\begin{eqnarray}
\label{formfactors}
&& \delta M =C_F \sum_{i=1}^{12} f_i M_i, \quad C_F = \frac{4}{3} \, ,
\end{eqnarray}
with the reduced matrix elements $M_i$ in~(\ref{me}) and 
form factors $f_i$, which are given explicitly in the Appendix.

Here we present the analytical results in the $\overline{MS}$ scheme;
accordingly, the quark masses in (\ref{Yukawa}) are running
masses $m_{b,t}(\mu)$.
The  results in the OS scheme can be easily obtained 
by using the pole masses and
replacing
the corresponding field- and mass-renormalization constants. 
Actually in the $\overline{MS}$ scheme, we must add the finite part of 
the wave-function renormalization constants for the $b$ quarks,
according to the LSZ prescription.

From the self-energy diagram in Fig.~1(h) for the $b,t$ quarks
we get the
explicit expressions of  the renormalization constants, valid
for both the $t$ and $b$ quark,
\begin{eqnarray}
 \frac{\delta m}{m} &=&
- \frac{\alpha_s}{4 \pi} 3 C_F \Delta,
\nonumber \\
 Z_L &=& Z_R = 
-\frac{\alpha_s}{4 \pi} C_F  \Delta, \quad \quad {\rm with}
\nonumber \\
\Delta &=& \frac{2}{\epsilon} -\gamma_E +\log(4\pi), \quad
\epsilon=4-D,
\end{eqnarray}
in dimensional regularization.

Squaring the amplitude and performing the spin summations,
\begin{eqnarray}
\overline{\sum} \, |M_{\rm ren}|^{2} =\overline{\sum} \, |M_0^{(s)}
+M_0^{(t)}|^{2} +2\, {\rm Re} \, \overline{\sum}\, \delta M (M_0^{(s)}
+M_0^{(t)})^{\dag} \, ,
\end{eqnarray}
yield the partonic
cross section for the process $b\bar b\rightarrow
W^{+}H^{-}$ with virtual corrections,
\begin{equation}
\label{virtual}
\hat{\sigma}^{\rm virt}_{b\bar{b}\rightarrow W^{+}H^{-}}(\hat{s})  = 
\int_{\hat{t}_{-}}^{\hat{t}_{+}}  d\hat{t} \, 
\frac{1}{16\pi\hat{s}^2} \, \overline{\Sigma} \, |M_{\rm ren}|^{2} 
\end{equation}
with
\begin{eqnarray}
\hat{t}_{\pm} &=& \frac{m_{W}^{2} +m_{H^{-}}^{2} -\hat{s}}{2} \pm
\frac{1}{2}\sqrt{(\hat{s} -(m_{W} +m_{H^{-}})^{2})(\hat{s} -(m_{W}
-m_{H^{-}})^{2})} \, .
\end{eqnarray}
The partonic cross section $\hat{\sigma}^{\rm virt}_{
b\bar{b}\rightarrow W^{-}H^{+}} $for the process 
$b\bar{b}\rightarrow W^{-}H^{+}$ is identical to that of 
$b\bar{b}\rightarrow W^{+}H^{-}$. 

\subsection{Real-gluon contributions}

The virtual corrections involve an infrared singularity from the
massless gluon. For our purpose, we can use a small gluon mass $\lambda$ 
for regularization, which allows to identify the infrared divergence
as a $\log(\lambda)$ term. 
This infrared divergence is cancelled by adding the corresponding
real-gluon-radiation corrections, displayed in Fig.~1(i-m). 
For technical reasons, it is convenient to perform the phase-space
integration over a soft- and hard-gluon part separately.
The soft-gluon contribution to the cross section 
at the partonic level is proportional
to the tree-level cross section 
$\hat \sigma^0_{b\bar b \rightarrow H^\pm
W^\mp}$ for $b\bar b \rightarrow H^\pm
W^\mp$,
\begin{eqnarray}
\hat \sigma^{\rm soft} &=&\hat \sigma^0_{b\bar{b}\rightarrow H^\pm
W^\mp} \,
\delta_s \, ,  \nonumber \\
\delta_s &=& - C_F \frac{\alpha_s}{2 \pi}
\{\log \frac{4 \Delta E^2}{\lambda^2}+
\log \frac{m_b^2}{\hat{s}}+
\log \frac{m_b^2}{\hat{s}} \log \frac{4 \Delta E^2}{\lambda^2}
+\frac{\pi^2}{3}+\frac{1}{2}  \log^2(\frac{m_b^2}{\hat{s}}) \},
\end{eqnarray}
where $\Delta E$ is the energy cutoff for soft gluons.
After adding the cross section (\ref{virtual}) with the
virtual corrections, the sum is independent on $\lambda$, 
and the $\log^2$ term cancels.

For the hard-gluon part,
we use the Monte Carlo packages BASES \cite{Bases} to perform
the  phase space integration, with the cutoff that
the energy of the gluon is greater than $\Delta E$. 
We do not give detailed  expressions here. 
Numerically it was  checked that the sum 
$\hat \sigma^{\rm soft} + \hat \sigma^{\rm hard}$ is independent 
of the cutoff  $\Delta E$.

At $O(\alpha_s)$, there are also initial-gluon contributions from
$gb \rightarrow b W^\pm H^\mp$, the Feynman diagrams can be obtained
from Fig.~1(i-m) by treating gluon and b quark as incoming
partons.  For the processes, we should caution how to subtract the
on-shell top quark and/or charged Higgs boson contributions.
Following the methods of Ref. \cite{Tait:2000cf}, the
partonic cross section 
after subtracting the on-shell contributions 
of subprocess  $gb \rightarrow b W^\pm H^\mp$ can
be written as:
\begin{eqnarray}
\hat \sigma^{0, sub}_{gb \rightarrow b W^\pm H^\mp}
&=&\hat \sigma^0_{gb\rightarrow b W^\pm H^\mp}
-\hat \sigma^0_{gb \rightarrow t H^-} Br(t \rightarrow b W^+)
\nonumber  \\
&&- \hat \sigma^0_{gb \rightarrow t W^-} Br( t
\rightarrow b H^+).
\end{eqnarray}
The cross section of $g \bar b \rightarrow \bar b  W^\pm H^\mp$ can
be obtained similarly.

\subsection{Hadronic cross section and subtracting double counting}

In the approach described above 
we have considered the real-gluon corrections, 
which give rise to a term involving 
$\log(\hat{s}/m_b^2)$ from the region where the
$b$ quark splits into a collinear $b$-quark--gluon pair and
the gluon into a collinear $b$ pair.
This logarithm is already contained in the heavy-quark distribution
function, hence it has to be removed. This can be done,
following~\cite{Dicus1},
by subtracting the tree-level process $b \bar b
\rightarrow H^\pm W^\mp$ convoluted  
with one heavy-quark distribution function
given by the perturbative solution to DGLAP equation,
\begin{eqnarray}
\tilde{b}_{(bb)}(x,\mu_f)&=& \frac{\alpha_s (\mu)}{2 \pi}
\log (\frac{ \mu_f^2}{m_b^2}) \int^{1}_{x}
\frac{d y}{y} P_{qq}(\frac{ x}{y}) b(y,\mu_f),
\nonumber \\
\tilde{b}_{(bg)}(x,\mu_f)&=& \frac{\alpha_s (\mu)}{2 \pi}
\log (\frac{ \mu_f^2}{m_b^2}) \int^{1}_{x}
\frac{d y}{y} P_{qg}(\frac{ x}{y}) g(y,\mu_f),
\end{eqnarray}
where
\begin{eqnarray}
P_{qq}(z)&=&C_F [\frac{1+z^2}{(1-z)_+}+\frac{3}{2}
\delta(1-z)],
\nonumber \\
P_{qg}(z)&=& \frac{1}{2} [z^2+(1-z)^2].
\end{eqnarray}
In this way, the total cross section at $O(\alpha_s)$  can be expressed
as
\begin{eqnarray}
\sigma^{NLO}&=& \bar{b}  \otimes \hat{\sigma}_{b\bar{b}}  \otimes b 
-\tilde{b}_{(bb)}  \otimes \hat \sigma^{0}_{
\bar b b \rightarrow H^\pm W^\mp} \otimes b
-\bar{b}  \otimes \hat \sigma^{0}_{
\bar b b \rightarrow H^\pm W^\mp} \otimes \tilde{b}_{(bb)}
 \nonumber \\
&&
+b \otimes \hat \sigma^{0, sub}_{
b g \rightarrow  b H^\pm W^\mp  } \otimes g
- b  \otimes \hat \sigma^{0}_{
\bar b b \rightarrow H^\pm W^\mp} \otimes \tilde{b}_{(bg)}
 \nonumber \\
&&+\bar b \otimes \hat \sigma^{0, sub}_{
\bar b g \rightarrow  \bar b H^\pm W^\mp  } \otimes g
- \bar b  \otimes \hat \sigma^{0}_{
\bar b b \rightarrow H^\pm W^\mp} \otimes \tilde{b}_{(bg)},
\label{he}
\end{eqnarray}
where 
$\hat{\sigma}_{b\bar{b}}$ is the infrared-finite parton cross section 
which is given by the sum of
$\hat \sigma^{\rm virt} + \hat \sigma^{\rm soft} + \hat \sigma^{\rm hard}$.
In eq. (\ref{he}),
$ A \otimes \hat \sigma \otimes B $ represents the cross section of the 
subprocess $\hat \sigma$ is convoluted with the parton distribution functions (PDF) $A$ and $B$,
\begin{equation}
A \otimes \hat \sigma \otimes B =\int_{z_0}^1  dz \,
\frac{dL}{dz}\,
\hat{\sigma} (z^2s)\, , \quad
z_0 = \frac{m_W +m_{H^-}}{\sqrt{s}} \, ,
\end{equation}
where $\sqrt{s}$ is the overall CM energy of the
$pp$ system,
and the parton luminosity $dL/dz$ is defined as
\begin{equation}
\frac{dL}{dz} =2z\int_{z^{2}}^{1}
\frac{dx}{x}\, A(x,\mu_f) \, B(\frac{z^2}{x},\mu_f).
\end{equation}

\section{Numerical results}

We now present a numerical discussion of the QCD corrections
to $W^{\pm}H^{\mp}$ associated production at the LHC. The SM input
parameters in our calculations were taken to be
$\alpha_{ew}(m_Z)=1/128.8$, $\alpha_s(m_Z)=0.118$,
$m_W=80.41$GeV and
$m_Z=91.1867$GeV, and the pole masses of
top and bottom quarks are $m_t=175$GeV and $m_b=4.7$GeV. We have
used the two-loop running $\overline{MS}$ quark masses \cite{quarkmass}
and strong
coupling constant \cite{PDG}, 
the CTEQ5M  PDF~\cite{CTEQ}, 
and we choose the factorization and
renormalization scale as $m_{H^\pm}+m_W$, if not stated otherwise.
The expressions given in the previous section are valid for a general
Two-Higgs-doublet model; they cover the 
MSSM case for a specific choice of
the Higgs-boson masses and mixing angles. 
Here we focus on the MSSM scenario, taking into account
the one-loop relations \cite{Mass} between the Higgs-boson masses
$M_{h,H,A,H^\mp}$ and the angles $\alpha$ and $\beta$, with 
$m_{H^+}$ and $\beta$  chosen as the two
independent Higgs-input parameters, together  with $M_S=1$ TeV 
as a genuine SUSY mass scale. As a remark, the
SUSY-QCD corrections arising from virtual gluino and squarks may
also become important for specific parameters, 
which will be studied separately.

In Fig.~2 we present the total cross sections,
for the $b\bar{b}$-annihilation subprocess,
in the $\overline{MS}$ and OS renormalization
schemes, and the relative deviation
$\Delta=(\sigma^{OS}-\sigma^{\overline{MS}})/(\sigma^{OS}+\sigma^{
\overline{MS}})$ in Fig.~3, 
both as functions of $m_{H^\pm}$. 
The results differ remarkably at tree-level (LO)  
and are significantly closer
at the level including the QCD corrections (NLO).
Whereas in the OS scheme the LO cross section is sizeably reduced
at NLO, the cross section increases in the \msbar\ scheme at NLO.

For $\tan\beta=50$, $\Delta$ is still relatively large, 
around $8\%$, which can, however,
be easily understood as a higher-order effect.
For large $\tan\beta$ 
we can write
the cross section in the two schemes as
\begin{eqnarray}
\sigma^{OS} & \propto & m_{b, pole}^2\, (1-A+B),  \nonumber \\
\sigma^{\overline{MS}} & \propto & m_b(\mu)^2\,  (1+B),
\end{eqnarray}
with
\begin{eqnarray}
 A = 2 \frac{\alpha_s}{\pi} (\log\frac{\mu^2}{m_b^2}+\frac{4}{3})\, ,
\end{eqnarray}
which is the quantity entering also the relation between 
the \msbar\ quark mass 
and the corresponding pole mass at one-loop order,
\begin{eqnarray}
 m(\mu)^2 =m_{pole}^2\,  
 [1-2\frac{\alpha_s}{\pi} (\log\frac{\mu^2}{m^2}+\frac{4}{3})] \, .
\end{eqnarray}
Hence, one finds 
\begin{eqnarray}
\Delta &\simeq &  A B/2 \, .
\end{eqnarray}
For $\mu=200 GeV$, we have $A \simeq 0.6$,  $B \simeq 0.3$  
and $\Delta \simeq 9\%$.
In the following we will give the results only in the $\overline{MS}$
scheme, which is usually more stable with respect to missing
higher-order terms.

Fig.~4 and Fig.~5 show the LO tree-level cross section $\sigma^0$
and the NLO cross section $\sigma^{NLO}$, 
including the $bg$ contribution, 
as well as the relative correction
$\delta=(\sigma^{NLO}-\sigma^0)/\sigma^0$
as a function of $m_{H^+}$ and $\tan\beta$.
As pointed out before, due to the
competition between the top--Higgs and bottom--Higgs 
Yukawa couplings,
the cross sections are relatively small for
intermediate values of $\tan\beta$,
around $\tan\beta \sim 6$ (see Fig.~6). 
From the figures, we can see that the QCD corrections are negative and the
magnitude is greater than $14\%$ for all charged Higgs mass and
$\tan\beta$.
For $\tan\beta=2$, the QCD corrections can decrease the cross section
more than $30\%$.
Table 1 contains numerical results for the relative correction $\delta$
for a low and a high value of $\tan\beta$,  
$\tan\beta=2$ and $50$.

Fig.~6 and 7 display the LO and NLO hadronic  cross section
and the relative correction $\delta$
as a function of
$\tan\beta$, for $m_{H^\pm}=200, 500$ and $1000$ GeV. 
From the figures, one can see that
the QCD corrections
are almost independent of
$\tan\beta$ when $\tan\beta>15$. 
For $\tan\beta<15$, 
the magnitude of the QCD corrections decreases with the increment of
$\tan\beta$.

In Fig. 8 we present the uncertainty of varying the 
factorization scale, which is defined as
\begin{eqnarray}
\frac{\sigma (\mu_f=2 \mu_0)-
\sigma (\mu_f=\frac{1}{2} \mu_0)}{\sigma (\mu_f=\mu_0)}
\label{un}
\end{eqnarray}
with $\mu_0=m_{H^\pm}+m_W$,
as a function of $m_{H^\pm}$. We can see that the
uncertainty is less than $10\%$ for all the charged Higgs  boson
mass and $\tan\beta$.
 

\section{Conclusion}

In conclusion, we have calculated the $O(\alpha_{s})$
corrections to the cross sections for $b \bar b \rightarrow 
W^{\pm}H^{\mp}$ associated
production at the LHC  in both $\overline{MS}$ and OS
renormalization schemes, with an explicit discussion for the
charged Higgs boson of the MSSM.
The results in
these two schemes are in good agreement.
Our numerical NLO results include also the contributios from
$gb\rightarrow b H^\pm W^\mp$.  
In the $\overline{MS}$ scheme,
the QCD corrections are negative and within $ -14\% \sim -20\%$ for charged
Higgs mass up to 1 TeV and 
$\tan\beta >15$. 
For $\tan\beta =2$,
 the magnitude of the QCD corrections can be greater than 
$30\%$. 

The analytical results given in this paper are also
valid in a general 2-Higgs-doublet model,
where the constrains among Higgs masses and angles $\alpha$
and $\beta$ are released. 
As a final remark, the higher-order contributions 
arising from the
$gg  \rightarrow H^\pm W^\mp b \bar b$,
may also be important, especially for lower values 
of the charged Higgs-boson  mass. 
They are presently under investigation.

\section{Acknowledgement}

S.H. Zhu  would like to thank Prof. C.S. Li
for the stimulating discussion
in the first stage of this work, and Prof. W.-K. Tung for
the suggestion on PDF.
This work was supported in part by the Alexander von Humboldt
Foundation,
National Nature Science Foundation of China and 
the Deutsche Forschungsgemeinschaft.
Parts of the calculations have been performed on the QCM
cluster at the University of Karlsruhe, supported by the
DFG-Forschergruppe ''Quantenfeldtheorie, Computeralgebra und
Monte-Carlo-Simulation''.

\newpage
\section{Appendix}

In this Appendix, we will present the non-vanishing form factors
in the decomposition~(\ref{formfactors}) of the various $s$- and $t$-channel
contributions to the one-loop matrix element.
The form factors contributing to $\delta M^{S(t)}$ are
\begin{eqnarray}
f_1&=& -\frac{\alpha_s g h_b m_b \sin\beta}{4 \sqrt{2} \pi
(m_t^2-\hat{t})^2 \hat{t}} [ (3 m_t^2-\hat{t})\hat{t}- m_t^2
(m_t^2+\hat{t}) B_0(0,0,m_t^2) \nonumber \\
&&+(m_t^4 - 6 m_t^2 \hat{t}+
\hat{t}^2)B_0(\hat{t},0,m_t^2)],\nonumber \\
f_2&=& -\frac{\alpha_s g h_t m_t \cos\beta}{2 \sqrt{2} \pi
(m_t^2-\hat{t})^2 } [ -m_t^2+ m_t^2 B_0(0,0,m_t^2)+(m_t^2 +
\hat{t})B_0(\hat{t},0,m_t^2)], \nonumber \\
f_3&=&f_1/m_b,\nonumber \\
f_9&=&2 f_1/m_b.
\end{eqnarray}

The form factors that contribute to $\delta M^{V_1(t)}$ are
(the argument the $C$ functions is $[m_b^2,\hat{t},m_w^2,m_b^2,
\lambda^2,m_t^2]$)
\begin{eqnarray}
f_1&=& \frac{\alpha_s g h_b m_b \sin\beta}{4 \sqrt{2} \pi
(m_t^2-\hat{t})} [ 1- 2 B_0(\hat{t},\lambda^2,m_t^2)-2
(m_b^2+m_t^2) C_0 - 2 m_w^2 (C_1+C_2) + 4 C_{00}], \nonumber \\
f_2&=& \frac{\alpha_s g h_t m_t \cos\beta}{4 \sqrt{2} \pi
(m_t^2-\hat{t})} [ -1+ 2 B_0(\hat{t},\lambda^2,m_t^2)+4 m_b^2 C_0
+ 2 (m_b^2+m_w^2-\hat{t}) C_1+ 2 m_w^2 C_2 - 4 C_{00}], \nonumber \\
f_3&=& \frac{\alpha_s g h_b \sin\beta}{4 \sqrt{2} \pi
(m_t^2-\hat{t})} [ 1- 2 B_0(\hat{t},\lambda^2,m_t^2)-2 m_b^2 C_0 -
2 (m_w^2- \hat{t})C_1- 2 m_w^2 C_2 + 4 C_{00}],\nonumber \\
f_4&=& \frac{\alpha_s g h_t m_t m_b \cos\beta}{2 \sqrt{2} \pi
(m_t^2-\hat{t})} [  C_0 +  C_1],\nonumber \\
f_9&=& - \frac{\alpha_s g h_b \sin\beta}{2 \sqrt{2} \pi
(m_t^2-\hat{t})} [ -1+ 2 B_0(\hat{t},\lambda^2,m_t^2)+2 m_b^2 C_0
- 2(m_b^2+m_t^2-m_w^2) C_1+ 2 m_w^2 C_2 \nonumber \\ && - 4 C_{00}
- 2 m_b^2 C_{11} -2 (m_b^2-\hat{t}) C_{12}],\nonumber \\
f_{10}&=& \frac{\alpha_s g h_t m_t m_b \cos\beta}{ \sqrt{2} \pi
(m_t^2-\hat{t})} [  C_0 +  2 C_1 +C_{11}],\nonumber \\
f_{11}&=& \frac{\alpha_s g h_b m_b \sin\beta}{\sqrt{2} \pi
(m_t^2-\hat{t})} [  C_1 +  C_{11}+C_{12}],\nonumber \\
f_{12}&=&- \frac{\alpha_s g h_t m_t \cos\beta C_{12}}{ \sqrt{2} \pi
(m_t^2-\hat{t})}.
\end{eqnarray}

The form factors that contribute to $\delta M^{box}$ are
(the argument of the $C$ and $D$ functions is $[m_b^2,m_b^2,\hat{s},m_b^2,
\lambda^2,m_b^2]$ and $[m_{H^{\pm}}^2,m_b^2,m_b^2,m_w^2,\hat{t},\hat{s},
m_t^2,m_b^2,\lambda^2,m_b^2]$)
\begin{eqnarray}
f_1 &=& \frac{\alpha_s g  m_b }{ 2 \sqrt{2} \pi} [ h_t m_b m_t
\cos\beta (D_0+D_2)+h_b \sin\beta (C_0 + m_t^2 
D_0+ m_{H^{\pm}}^2  D_1 \nonumber \\ &&+ 
(
 m_{H^{\pm}}^2 +m_w^2 -\hat{u}) D_2 - (m_b^2-\hat{u})D_3)],\nonumber \\ 
f_2 &=& \frac{\alpha_s g  }{ 2 \sqrt{2} \pi} [ h_t m_t \cos\beta
(-C_0+  ( m_b^2 -m_t^2 - \hat{u}) D_0 +
 (m_b^2- m_{H^{\pm}}^2 -\hat{u}) D_1 \nonumber \\
&&+ (m_b^2- m_{H^{\pm}}^2 -m_w^2) D_2 + 
(m_b^2- m_{w}^2 -\hat{u}) D_3+ 2 
D_{00}) \nonumber
\\ &&
 +h_b m_b \sin\beta
(-C_0 -  m_t^2 D_0 + (m_b^2-
\hat{u}) D_1 - \hat{t} D_2 \nonumber \\ && - 
 m_w^2  D_3+ 2  D_{00})],\nonumber \\ 
f_3 &=& \frac{\alpha_s g  }{ 2 \sqrt{2} \pi} [ h_t m_b m_t
\cos\beta (D_0+D_1+D_2+D_3) + h_b \sin\beta (m_b^2 (D_1+D_3) -
(2 m_b^2-\hat{s}) D_2)],\nonumber \\ 
f_4 &=& -\frac{\alpha_s g m_b }{ 2 \sqrt{2} \pi} [ h_t m_t
\cos\beta (D_0+D_1+D_2+D_3) + h_b m_b \sin\beta (D_1+D_3)],\nonumber \\ 
f_5 &=& \frac{\alpha_s g }{  \sqrt{2} \pi} [ h_t m_b m_t \cos\beta
D_{11}+  h_b \sin\beta (2 C_0 + m_t^2  D_0+ 
(C_1+C_2) \nonumber \\ && -  m_t^2  D_1 +
 \hat{t}
  D_2+m_w^2   D_3-2  D_{00}+   
  (m_b^2-2 m_{H^{\pm}}^2)
  D_{11}\nonumber \\
  && + (m_b^2-m_{H^{\pm}}^2-\hat{t})D_{12}+
  (2 m_b^2-\hat{t} -\hat{u}) D_{13})],\nonumber \\ 
f_6 &=& -\frac{\alpha_s g m_b }{  \sqrt{2} \pi} [ h_t m_t
\cos\beta (D_0+2 D_1+D_2+D_{11}+D_{12}) + h_b m_b \sin\beta (D_1+
D_{11}+D_{12})],\nonumber \\ 
f_7 &=& \frac{\alpha_s g h_b m_b D_3 \sin\beta}{  \sqrt{2} \pi},\nonumber \\ 
f_8 &=& -\frac{\alpha_s g }{  \sqrt{2} \pi} [ h_t m_t
\cos\beta (D_0 +2 D_1+D_2+D_3+D_{11}+D_{12} +D_{13}) \nonumber
\\&&+ h_b m_b \sin\beta (D_1+D_{11}+ D_{12}+D_{13})],\nonumber \\ 
f_9 &=& \frac{\alpha_s g  }{  \sqrt{2} \pi} [ h_t m_b m_t
\cos\beta (D_{11}+D_{12})+  h_b \sin\beta ( 2 C_0+  m_t^2
(D_0-D_1) \nonumber \\ && +  C_2+ 
(m_b^2-m_t^2+m_w^2-\hat{u})
  D_2 + m_w^2  D_3- 2  D_{00}\nonumber \\ &&
 + (m_b^2-
 2 m_{H^{\pm}}^2) 
 D_{11}+  (2 m_b^2- 3
 m_{H^{\pm}}^2-\hat{t}) 
 D_{12} \nonumber \\ &&+ (2 m_b^2-\hat{t}-\hat{u}) 
 D_{13}+ (m_b^2-m_{H^{\pm}}^2-\hat{t}) 
 D_{22} \nonumber \\ &&  + (2 m_b^2-\hat{t}-\hat{u}) 
 D_{23})],\nonumber \\ 
f_{10} &=& -\frac{\alpha_s g m_b }{  \sqrt{2} \pi} [ h_t m_t
\cos\beta (D_0+2 D_1 + 2 D_2+D_{11}+2 D_{12}+D_{22}) \nonumber \\
&& + \sin\beta h_b m_b (D_1+D_2+D_{11}+2 D_{12}+D_{22})],\nonumber \\ 
f_{12} &=& -\frac{\alpha_s g }{  \sqrt{2} \pi} [ h_t m_t \cos\beta
(D_1+D_2+D_{11}+2 D_{12}+D_{13}+D_{22}+D_{23}) \nonumber \\ && h_b
m_b \sin\beta (D_2+D_{11}+2 D_{12}+D_{13}+D_{22}+D_{23})].
\end{eqnarray}

The form factors  contributing to  $\delta M^{V_1(s)}(A)$ are
(the argument of the $C$ functions is $[m_b^2,m_b^2,\hat{s},m_b^2,
\lambda^2,m_b^2]$)
\begin{eqnarray}
f_5 &=& f_9=\frac{\alpha_s g h_b \sin\beta}{ 2 \sqrt{2} \pi
(\hat{s}-m_{A0}^2)} [-1+ 2 B_0(m_b^2,m_b^2,\lambda^2) +2 m_b^2
C_0+\hat{s} C_1+ 2 \hat{s} C_2], \nonumber \\
f_6&=&f_{10}=-f_5.
\end{eqnarray}

The form factors which contribute to  $\delta M^{V_1(s)}(H)$ are
(the argument of the $C$ functions is $[m_b^2,m_b^2,\hat{s},m_b^2,
\lambda^2,m_b^2]$, the form factors to  $\delta M^{V_1(s)}(h)$
can be obtained by substituting 
 $\cos\alpha \sin(\alpha-\beta) \, \rightarrow \,
  - \sin\alpha \cos(\alpha-\beta)$)
\begin{eqnarray}
f_5 &=& f_6=f_9=f_{10}=\frac{\alpha_s g h_b \cos\alpha
\sin(\alpha-\beta)}{ 2 \sqrt{2} \pi (\hat{s}-m_{H}^2)} [1- 2
B_0(m_b^2,m_b^2,\lambda^2) \nonumber \\ && -4 m_b^2 C_0 -(2 m_b^2
+ \hat{s}) C_1- 2 \hat{s} C_2].
\end{eqnarray}

The form factors which contribute to  $\delta M^{V_2(t)}$ are 
(the argument of the  $C$ functions is $[\hat{t},m_b^2, m_{H^{\pm}}^2,m_t^2,
\lambda^2,m_b^2]$)
\begin{eqnarray}
f_1 &=&\frac{\alpha_s g m_b }{ 2 \sqrt{2} \pi (\hat{t}-m_{t}^2)} [
h_t m_b m_t \cos\beta (C_0+C_1)+ h_b \sin\beta (-1 +2
B_0(m_b^2,m_b^2,\lambda^2)\nonumber \\ &&+3 m_t^2
C_0+(-m_b^2+m_{H^{\pm}}^2+ m_t^2+\hat{t}) C_1 +(-m_b^2+2
m_{H^{\pm}}^2+ m_t^2) C_2)],\nonumber \\ 
f_2 &=&\frac{\alpha_s g }{ 2 \sqrt{2} \pi (\hat{t}-m_{t}^2)} [ h_t
m_t \cos\beta(1-2 B_0(m_b^2,m_b^2,\lambda^2)-(2 m_t^2+\hat{t}) C_0
\nonumber \\ &&+(m_b^2- m_{H^{\pm}}^2- 2 \hat{t} )C_1 + (m_b^2-2
m_{H^{\pm}}^2 - \hat{t}) C_2) + h_b m_b \sin\beta (-m_t^2 C_0
\nonumber \\ &&  -\hat{t}C_1 + (m_t^2-\hat{t})C_2)],\nonumber \\ 
f_9&=&2 f_3 =2 f_1/m_b.
\end{eqnarray}

The scalar integrates $B,C,D$ in above expressions
are evaluated by FF package \cite{vanOldenborgh:1990wn}.

\newpage

\begin{figure}
\epsfxsize=15 cm
\centerline{\epsffile{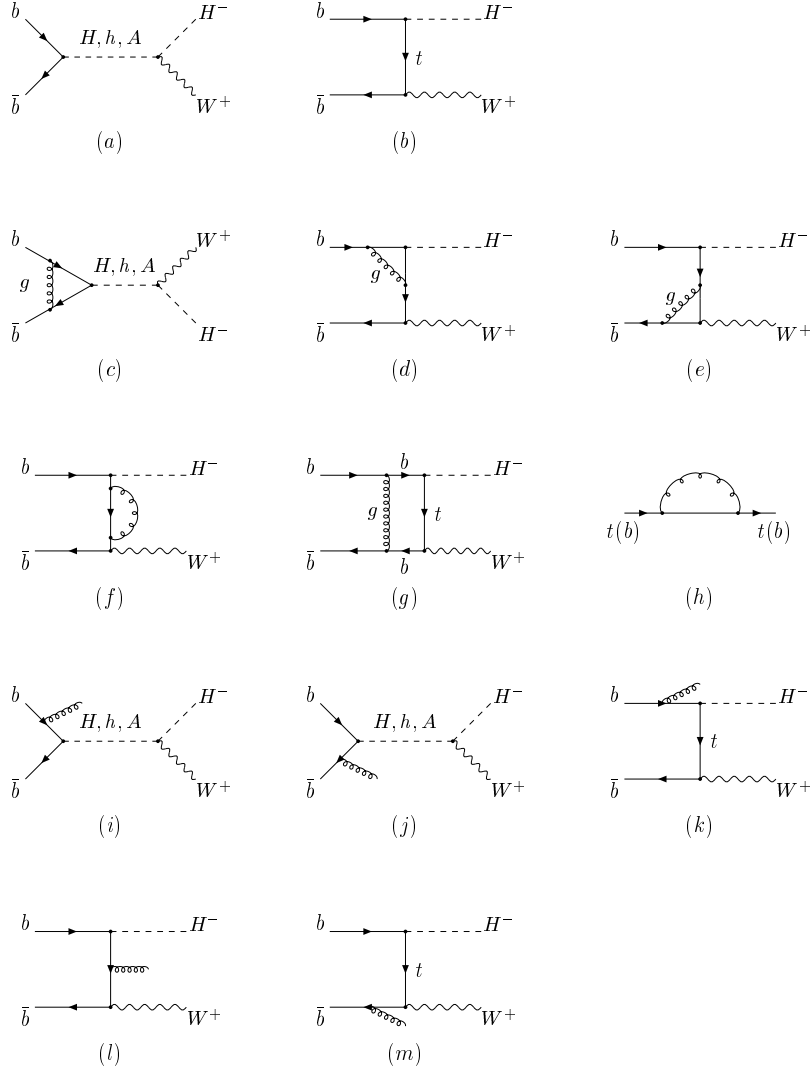}}
\caption[]{Feynmann diagrams for the subprocess $  b \bar b \rightarrow 
H^{-} W^+$ at the NLO: (a)-(b) Born diagrams;
(c)-(h) virtual correction diagrams; (i)-(m) gluon-radiation diagrams.}
\end{figure}

\newpage

\begin{figure}
\epsfxsize=13 cm 
\centerline{\epsffile{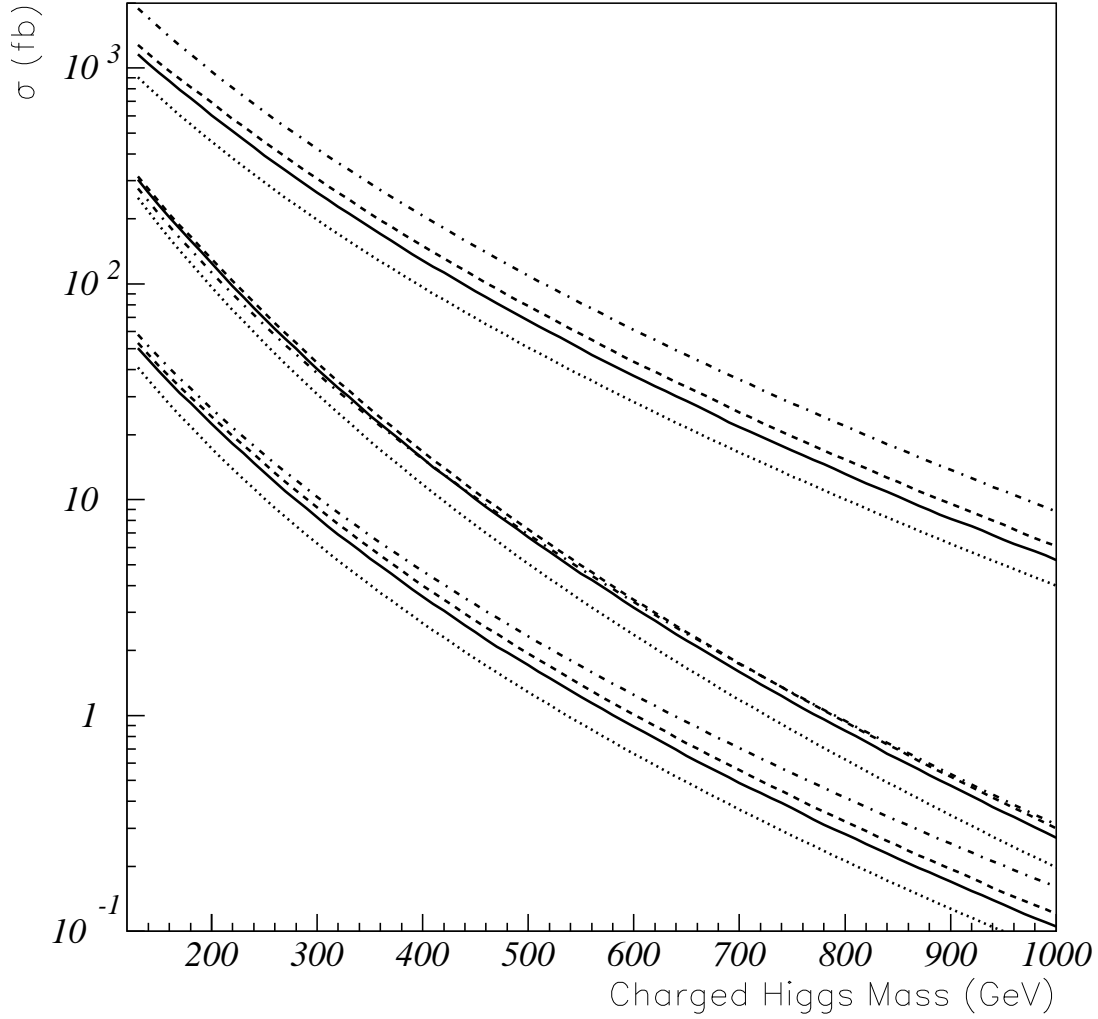}} \caption[]{ 
The total cross sections of process 
$PP \rightarrow b \bar b \rightarrow
H^{\pm} W^\mp$
versus $m_{H^\pm}$ at the LHC with $\sqrt{s}=14$
TeV in $\overline{MS}$ and OS schemes. 
The lines  correspond to the 
 cross sections at the
one-loop level of $\overline{MS}$ (solid),
one-loop level of OS (dashed), 
tree level of  $\overline{MS}$ (dotted) and 
tree level of OS (dash-dotted).
The three groups correspond to $\tan\beta=50, 2, 6$ from above to
the bottom.}
\end{figure}

\begin{figure}
\epsfxsize=13 cm 
\centerline{\epsffile{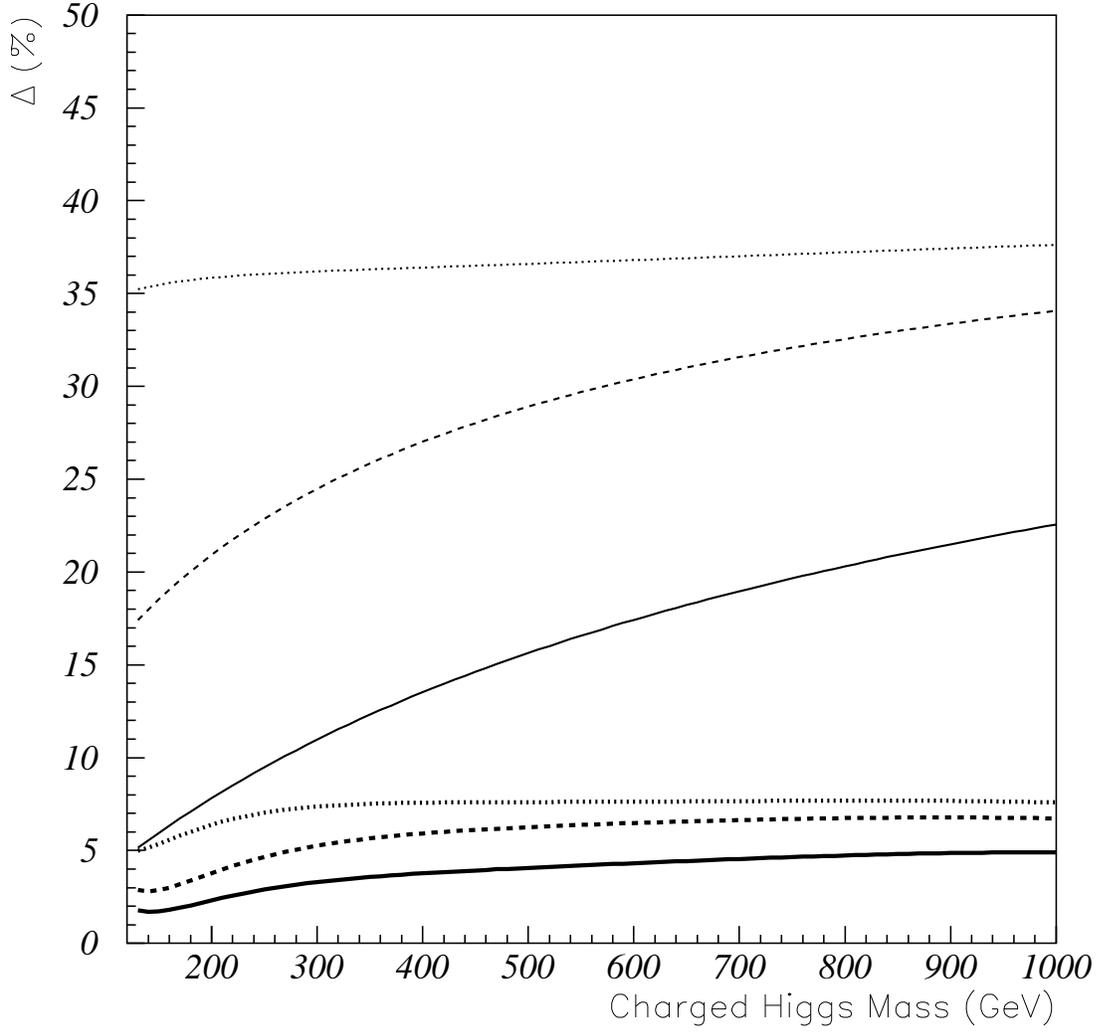}} \caption[]{ 
The $\Delta$ (defined in text)
versus $m_{H^\pm}$. The thick lines stand for the results
of one-loop and thin lines  stand for the results
of tree level. 
$\tan\beta= 2$ (solid), 6 (dashed), 50 (dotted).  }
\end{figure}

\begin{figure}
\epsfxsize=13 cm 
\centerline{\epsffile{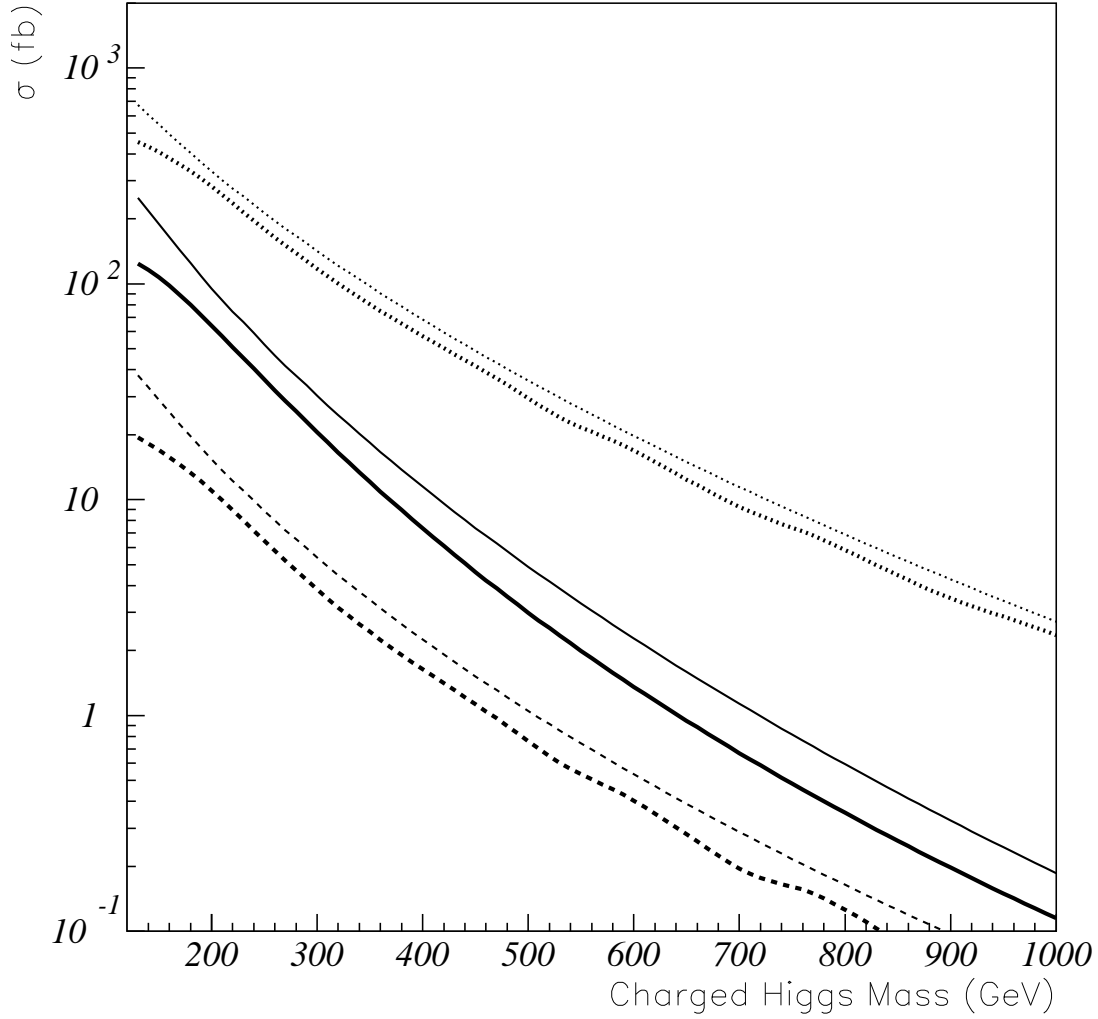}} \caption[]{ 
The
 total cross sections for the subprocesses 
$b \bar b \rightarrow
H^{\pm} W^\mp$ and $gb (\bar b) \rightarrow
H^{\pm} W^\mp b (\bar b)$
versus $m_{H^\pm}$ at the LHC with $\sqrt{s}=14$
TeV. The thick lines correspond to
the one-loop cross sections with
$\tan\beta=2$ (solid), $6$(dashed) and $50$ (dotted). The tree
level cross sections  are also shown
by the corresponding thin lines.
}
\label{FIG1}
\end{figure}

\begin{figure}
\epsfxsize=13 cm \centerline{\epsffile{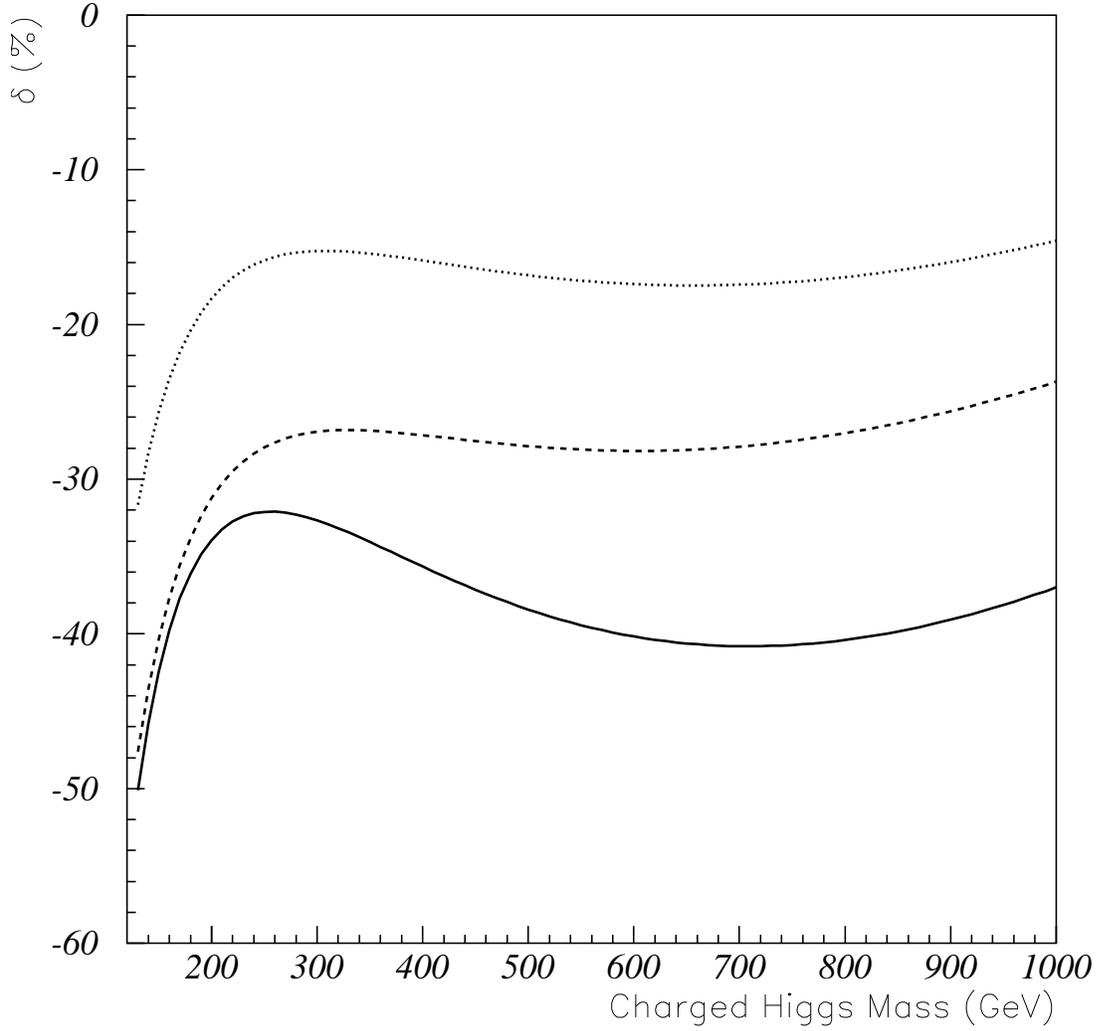}} \caption[]{ 
The QCD relative corrections to the cross sections for
the subprocesses
$b \bar b \rightarrow
H^{\pm} W^\mp$ and $gb (\bar b) \rightarrow
H^{\pm} W^\mp b (\bar b)$
versus $m_{H^\pm}$. 
The lines correspond to
$\tan\beta=2$(solid), $6$(dashed) and $50$(dotted). 
}
\label{FIG2}
\end{figure}

\begin{figure}
\epsfxsize=13 cm \centerline{\epsffile{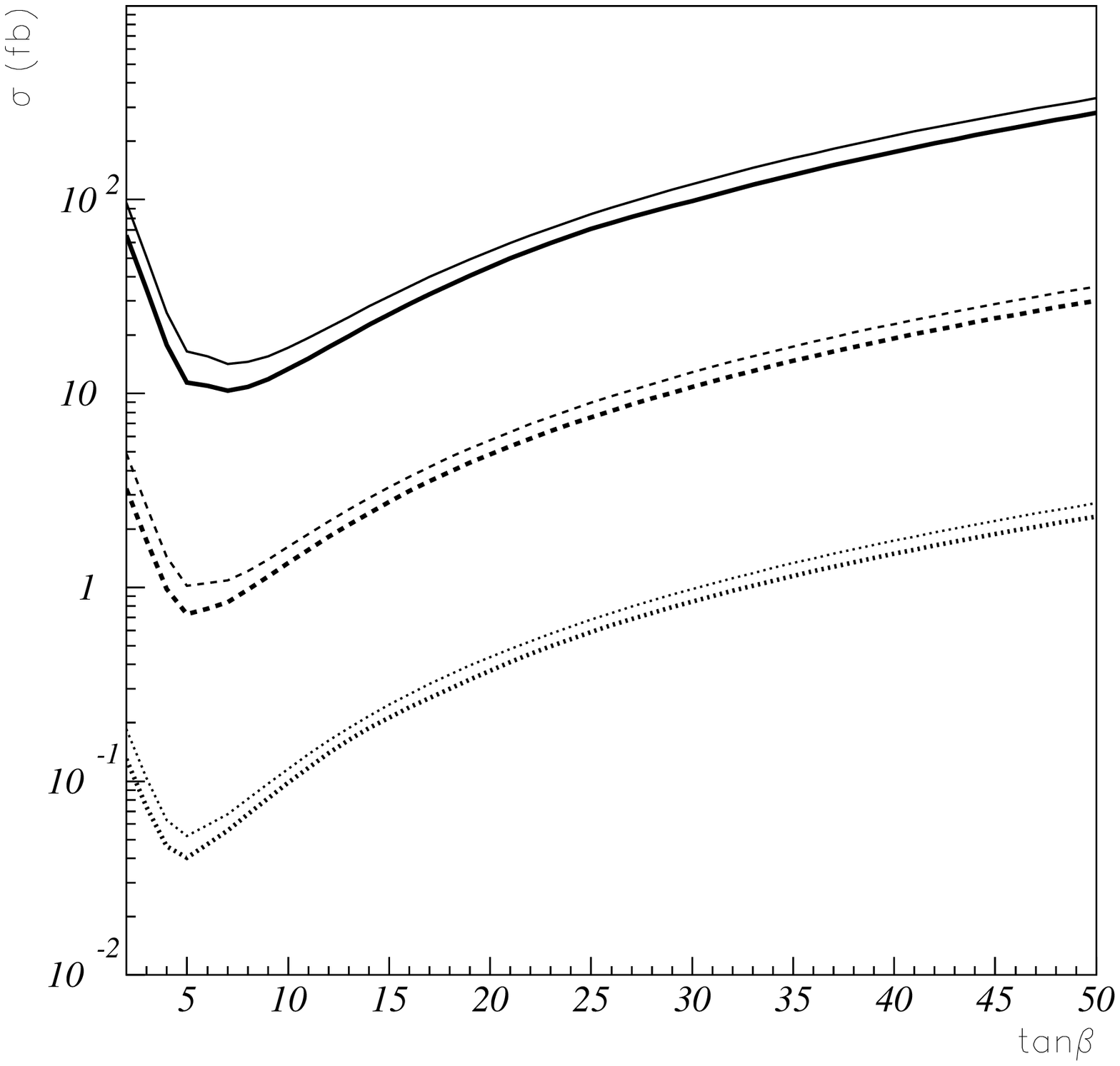}} 
\caption[]{ 
Same as Fig. \ref{FIG1} except
versus $\tan\beta$.
The lines correspond to
$m_{H^\pm}=200$(solid), $500$(dashed) and $1000$ GeV (dotted). 
}
\end{figure}

\begin{figure}
\epsfxsize=13 cm \centerline{\epsffile{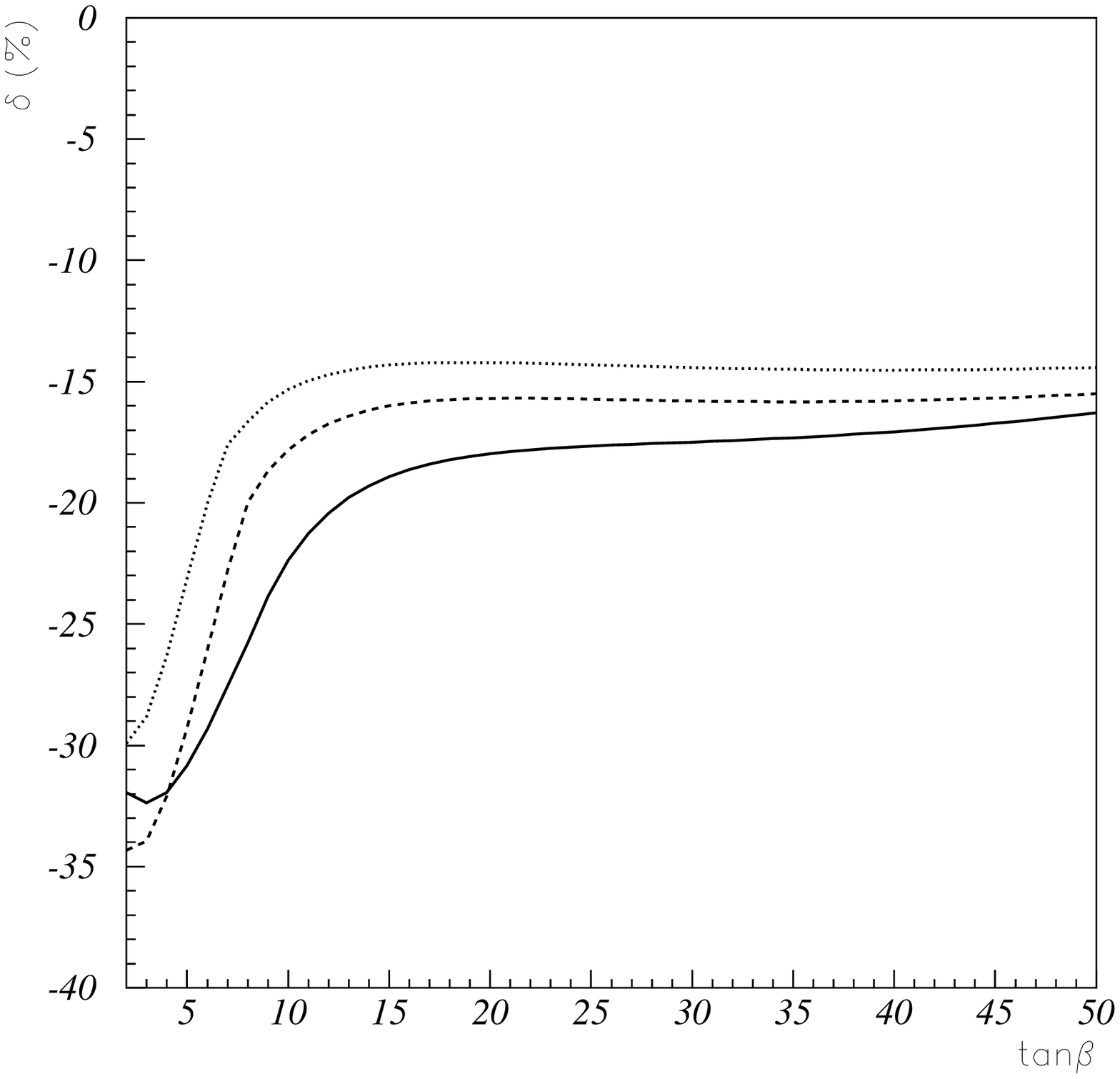}} \caption[]{ 
Same as Fig. \ref{FIG2} except 
versus $\tan\beta$. 
The lines correspond to
$m_{H^\pm}=200$(solid), $500$(dashed) and $1000$ GeV (dotted). 
}
\end{figure}

\begin{figure}
\epsfxsize=13 cm \centerline{\epsffile{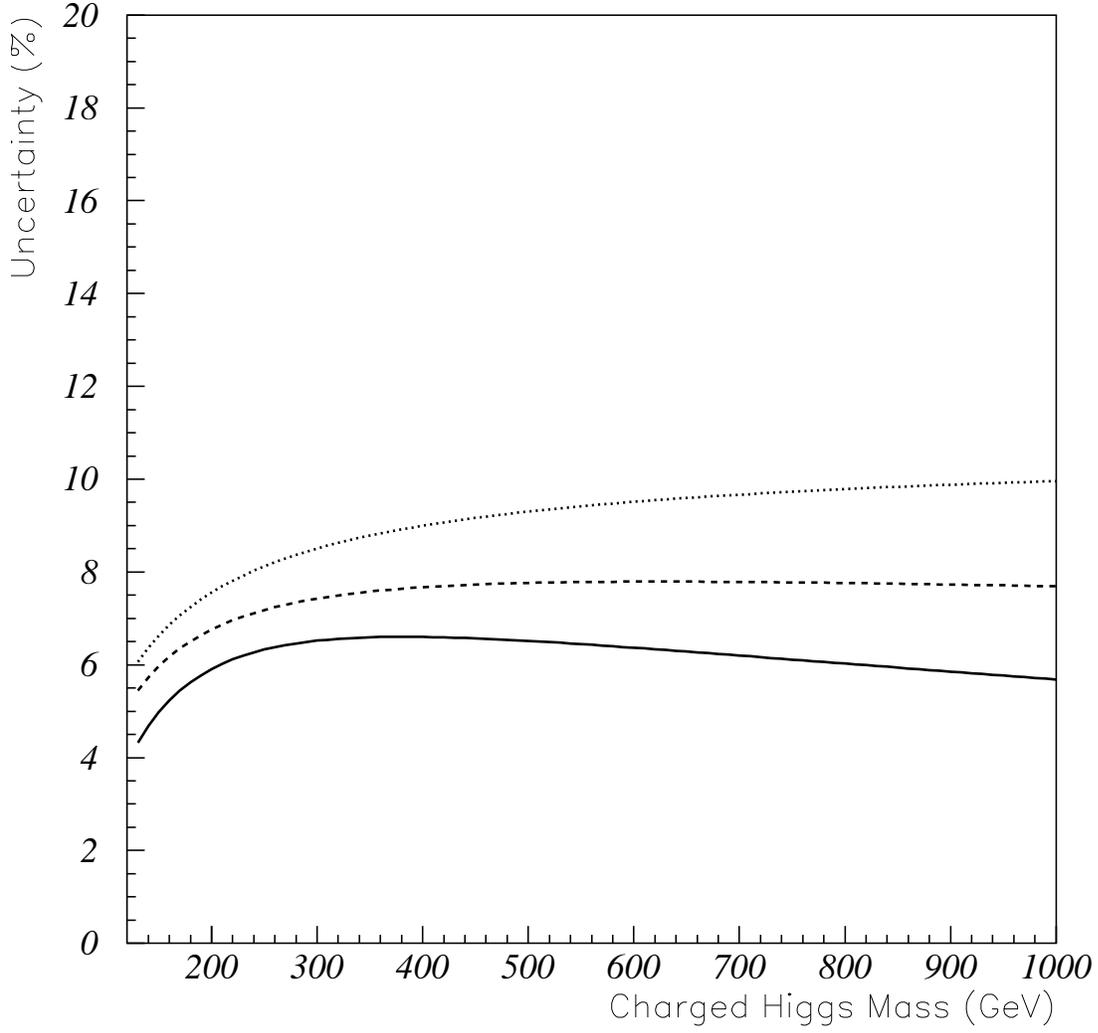}} \caption[]{ 
The uncertainty [defined in Eq. (\ref{un}) ]  as 
a function of $m_{H^\pm}$, where the renormalization scale
is fixed as $m_{H^\pm}+m_W$.
The lines correspond to
$\tan\beta=2$(solid), $6$(dashed) and $50$(dotted).
}
\end{figure}

\newpage
\begin{table}[t]
\begin{center}
\begin{tabular}{ccc}
\hline
$m_{H^\pm}$ (GeV) &  $\tan\beta =50$ & $\tan\beta=2$ \\
\hline
200  & $\sim$ -16\%  & $\sim$ -32 \% \\
\hline
500 &  $\sim$ -15\%  &  $\sim$ -34 \%\\
\hline
1000 &  $\sim$ -14\%     &$\sim$ -30 \%\\
\hline
\end{tabular}
\caption{QCD corrections to the process  $b\bar b \rightarrow
H^\pm W^\mp$ in MSSM and 2HDM in $\overline{MS}$ scheme. 
The renormalization and factorization scales are taken as 
$m_{H^\pm}+m_W$. } 
\end{center}
\end{table}

\end{document}